\pdfoutput=1
\documentclass[11pt]{article}
\usepackage[final]{acl}
\usepackage{times}
\usepackage{latexsym}
\usepackage[T1]{fontenc}
\usepackage[utf8]{inputenc}
\usepackage{microtype}
\usepackage{inconsolata}
\usepackage{amsmath}
\usepackage{amsfonts}
\usepackage{hyperref}
\usepackage{url}
\usepackage{tikz}
\usepackage{enumitem}
\usepackage[most]{tcolorbox}
\usepackage{svg}
 \usepackage{multirow} 
\usepackage[edges]{forest}
\definecolor{hidden-draw}{RGB}{0,0,0}
\usepackage{adjustbox}

\title{Negative Sampling Techniques in Information Retrieval: A Survey}

\author{
  \textbf{Laurin Wischounig\thanks{These authors contributed equally.}, 
  Abdelrahman Abdallah\footnotemark[1], Adam Jatowt} \\
  University of Innsbruck \\
  \texttt{\{abdelrahman.abdallah,adam.jatowt\}@uibk.ac.at} \\
  \texttt{laurin.wischounig@student.uibk.ac.at}
}

\begin{document}
\maketitle

\begin{abstract}
Information Retrieval (IR) is fundamental to many modern NLP applications. The rise of dense retrieval (DR), using neural networks to learn semantic vector representations, has significantly advanced IR performance. Central to training effective dense retrievers through contrastive learning is the selection of informative negative samples. Synthesizing 35 seminal papers, this survey provides a comprehensive and up-to-date overview of negative sampling techniques in dense IR. Our unique contribution is the focus on modern NLP applications and the inclusion of recent Large Language Model (LLM)-driven methods, an area absent in prior reviews. We propose a taxonomy that categorizes techniques including random, static/dynamically mined, and synthetic datasets. We then analyze these approaches with respect to trade-offs between effectiveness, computational cost, and implementation difficulty. The survey concludes by outlining current challenges and promising future directions for the use of LLM-generated synthetic data.
\end{abstract}

\section{Introduction}

Information Retrieval (IR)~\cite{baeza1999modern,bajaj2016ms,gruber-etal-2025-complextempqa,thakur2021beir} is a foundational field concerned with finding relevant information, typically within large collections of unstructured data (like text documents), that satisfies a user's or system's information need, often expressed as a query. 
IR powers a multitude of downstream applications including web search, QA and retrieval-augmented generation (RAG).

Traditionally, IR systems relied on lexical sparse retrieval methods, such as Term Frequency-Inverse Document Frequency (TF-IDF) like BM25~\cite{robertson2009probabilistic}. These techniques while effective and efficient, often struggle with semantic understanding, vocabulary mismatch (synonyms, paraphrasing), and capturing deeper contextual relationships.
In dense retrieval (DR)~\cite{zhao2024dense,chen2021salient,karpukhin2020dense,abdallah2025tempretriever,sciavolino2021simple}, queries and documents are encoded into relatively low-dimensional, dense vector representations (embeddings) using neural networks. The relevance score is typically computed based on the similarity (e.g., dot product or cosine similarity) between the query embedding and document embedding in this shared semantic space. Dense retrieval excels at capturing semantic meaning beyond keyword matching. Learning effective dense representations is paramount, and contrastive learning has emerged as a dominant paradigm. This approach trains a model to pull relevant query-document pairs closer together in the embedding space while pushing irrelevant pairs apart.

The strategic selection of these "irrelevant" pairs, referred to as negative sampling~\cite{yang2024does,abdallah-etal-2025-dear,abdallah-etal-2025-good}, is a decisive factor in a model's final performance. The evolution of negative sampling strategies directly maps to performance gains on benchmarks like MS MARCO~\cite{bajaj2016ms,lin2021batch}: while random in-batch negatives yield a baseline MRR@10 of $0.261$, incorporating static "hard" negatives from BM25 improves it to $0.299$. A further leap to 0.330 MRR@10 was achieved by ANCE~\cite{Xiong2020ANCE}, which introduced dynamic mining to ensure the model is continuously challenged by difficult negatives it finds for itself. This relentless pursuit of harder negatives, however, revealed the false negative problem. The very process of mining top-ranked documents for negatives risks including genuinely relevant but unlabeled passages. RocketQA~\cite{Qu2021RocketQA} quantified this risk, estimating that 70\% of top-retrieved but unlabeled passages are actually positives, a contamination that can poison the training data and severely degrade model performance. This turned out to be such a big problem, that for a long time a lot of research effort on negative sampling in NLP focused on false negative mitigation.

This survey aims to provide a structured and contemporary overview of this critical area. 
While other reviews exist~\cite{Xu2022Review,Yang2024TPAMI}, they neither focus on negative sampling for dense retrieval nor capture the recent, transformative impact of Large Language Models (LLMs) on negative sampling~\cite{zhao2025can}. Our unique contribution is to synthesize the relevant literature with a specific focus on dense retrieval for NLP applications~\cite{li2024syneg}, by categorizing 35 seminal papers. Also, we focus specifically on negative sampling techniques for contrastive learning in dense retrieval. While we acknowledge complementary approaches like ColBERT's late interaction mechanisms \cite{Khattab2020ColBERT} and knowledge distillation \cite{Lin2021RepL4NLP}, these operate at different levels (architecture vs. sampling strategy) and are appropriately positioned as orthogonal techniques that can be combined with any negative sampling approach discussed here.

The survey addresses three key research questions: \textbf{(RQ1)} How can the diverse landscape of negative sampling techniques for dense retrieval be categorized, and what are the core principles, advantages, and trade-offs of each approach? \textbf{(RQ2)} How can the diverse landscape of negative sampling techniques for dense retrieval be categorized, and what are the core principles, advantages, and trade-offs of each approach? \textbf{(RQ3)} What are the emerging directions and future challenges for negative sampling, particularly with the rise of generative and data-centric methods powered by LLMs?

\section{Related Work}

General reviews of negative sampling exist for machine learning \citep{Xu2022Review} and dense retrieval \citep{Yang2024TPAMI}, but neither focuses on dense retrieval methods that have become central to modern NLP applications such as retrieval-augmented generation (RAG), question answering, and dialogue systems. The LLM4IR~\cite{LLM4IRSurvey} survey provides broad coverage of LLM applications across IR components, but does not provide a systematic taxonomy focused on negative sampling for contrastive learning, nor does it analyze the critical false negative problem and mitigation strategies that are central to our review. Our survey provides the first comprehensive framework specifically examining negative sampling techniques for dense retrieval in NLP contexts, with emphasis on recent LLM-driven methods absent from prior reviews. We focus exclusively on negative sampling for contrastive learning; complementary approaches like knowledge distillation \citep{Lin2021RepL4NLP}, inference-time query augmentation \citep{gao2022hyde,abdallah2025asrank,abdallah2025dynrank}, late interaction architectures \citep{Khattab2020ColBERT}, and refined representations \citep{ji2025learningeffectiverepresentationsdense} are noted but not surveyed in depth. For extended discussion of related work and positioning relative to existing surveys, we refer the reader to Appendix~\ref{app:related_work}.

\section{Contrastive Learning for Dense Representations}

The core idea behind contrastive learning for dense retrieval is to train encoder models $E_q$ (for queries) and $E_p$ (for passages/documents) such that the similarity score $\text{sim}(E_q(q), E_p(p))$ is high for relevant pairs $(q, p^+)$ and low for irrelevant pairs $(q, p^-)$, commonly referred to as triplet loss. Architectures like Sentence-BERT \citep{Reimers2019SentenceBERT} are commonly trained using such contrastive objectives.
A common objective function used is based on Noise Contrastive Estimation (NCE), often implemented as the InfoNCE loss \citep{oord2019_infonce}. Given a query $q$, its positive (relevant) passage $p^+$, and a set of negative (irrelevant) passages $\{p_i^-\}_{i=1}^N$, let $S = \{p^+\} \cup \{p_i^-\}_{i=1}^N$ be the set of all candidate passages for query $q$. The goal is to minimize the negative log-likelihood of correctly identifying the positive passage among the candidates:

\begin{equation}
\begin{split}
&L(q, p^+, S) = \\ &-\log \frac{\exp(\text{sim}(E_q(q), E_p(p^+))/\tau)}{\sum_{p \in S} \exp(\text{sim}(E_q(q), E_p(p))/\tau)}
\end{split}
\end{equation}

Here, $\text{sim}(\cdot,\cdot)$ is a similarity function (e.g., dot product), and $\tau$ is a temperature hyperparameter scaling the similarities~\cite{manna2025dynamically}. The effectiveness of this learning process hinges on the choice of the negative samples $\{p_i^-\}$. If negatives are too "easy" (semantically very distant from the query), the model receives weak learning signals and may fail to distinguish between relevant passages and challenging irrelevant ones during inference. Conversely, selecting informative "hard" negatives forces the model to learn finer-grained semantic distinctions, leading to more robust and accurate representations \citep{Zhan2021HardNegatives}.

\section{Taxonomy of Negative Sampling Techniques}
\label{sec:taxonomy}
Since dense representation learning is a multifaceted problem, where many aspects of a model can be improved, researchers have proposed many different techniques~\cite{zhan2020learning,lindgren2021efficient,zhan2021optimizing,li2024domain}. The common objective optimized for are the metrics mentioned above used for quantifying the retrieval performance such as precision, recall and nDCG@k, but also performance on downstream tasks. On the other hand, negative sampling techniques are also often used to improve training efficiency, training stability, sample efficiency, generalizability and domain transferability.

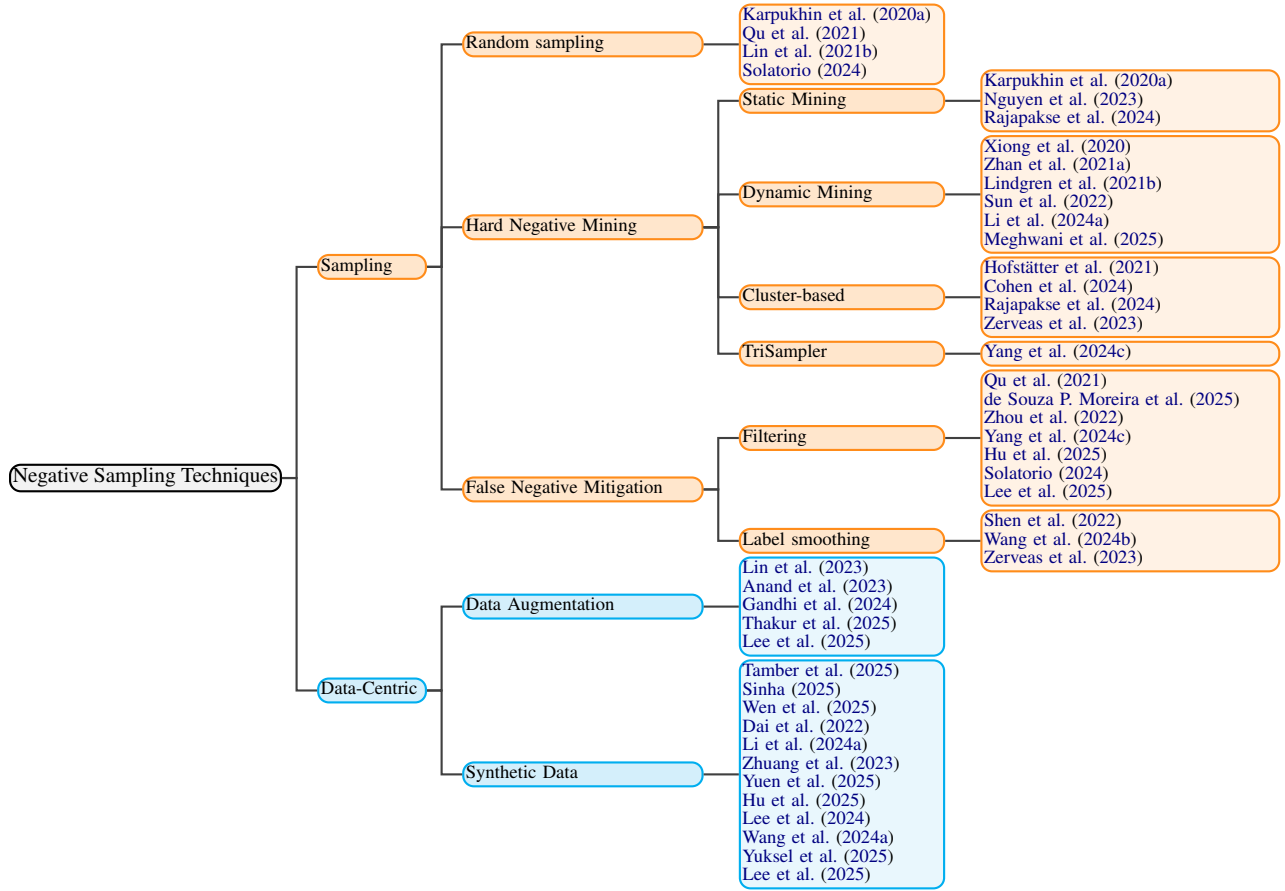
\begin{figure*}[ht]
    \centering
    \tikzstyle{my-box}=[
        rectangle,
        rounded corners,
        minimum height=0.1em,
        minimum width=0.1em,
        inner sep=1pt,
        align=left,
        fill opacity=.5,
        text opacity=1,
    ]
    \tikzstyle{leaf}=[my-box]
    \begin{forest}
        forked edges,
        for tree={
            grow=east,
            reversed=true,
            anchor=base west,
            parent anchor=east,
            child anchor=west,
            base=left,
            font=\fontsize{8}{8}\selectfont,
            rectangle,
            rounded corners,
            align=left,
            minimum width=0.1em,
            edge+={darkgray, line width=0.8pt},
            s sep=1pt,
            inner xsep=1pt,
            inner ysep=1pt,
            draw=hidden-draw,
            ver/.style={rotate=90, child anchor=north, parent anchor=south, anchor=center},
        },
        where level=1{text width=3.5em,font=\fontsize{7}{7}\selectfont,}{},
        where level=2{text width=8em,font=\fontsize{7}{7}\selectfont,}{},
        where level=3{text width=6.8em,font=\fontsize{7}{7}\selectfont,}{},
        where level=4{text width=10em,font=\fontsize{7}{7}\selectfont,}{},
        where level=5{text width=6.8em,font=\fontsize{7}{7}\selectfont,}{},
        [
            Negative Sampling Techniques, draw=black, fill=gray!10, thick, text=black,
            [
                Sampling,
                color=orange!90, fill=orange!20, thick, text=black
                [
                    Random sampling,
                    color=orange!90, fill=orange!20, thick, text=black
                    [
                        \citet{Karpukhin2020DPR}\\
                        \citet{Qu2021RocketQA}\\
                        \citet{Lin2021RepL4NLP}\\
                        \citet{solatorio2024gistembedguidedinsampleselection}\\
                        ,color=orange!90, fill=orange!20, thick, text=black, leaf
                    ]
                ]
                [
                    Hard Negative Mining,  
                    color=orange!90, fill=orange!20, thick, text=black
                    [
                        Static Mining,
                        color=orange!90, fill=orange!20, thick, text=black
                        [
                            \citet{Karpukhin2020DPR}\\
                            \citet{Nguyen2023PassageBM25}\\ 
                            \citet{rajapakse2024negativesamplingtechniquesmultilingualsetting}
                            ,color=orange!90, fill=orange!20, thick, text=black, leaf
                        ]
                    ]
                    [
                        Dynamic Mining,
                        color=orange!90, fill=orange!20, thick, text=black
                        [
                            \citet{Xiong2020ANCE}\\
                            \citet{Zhan2021HardNegatives}\\
                            \citet{Lindgren2021Cache}\\
                            \citet{sun2022reducecatastrophicforgettingdense}\\
                            \citet{li2024conanembeddinggeneraltextembedding}\\
                            \citet{meghwani2025hardnegativeminingdomainspecific}\\
                            ,color=orange!90, fill=orange!20, thick, text=black, leaf
                        ]
                    ]
                    [
                        Cluster-based,
                        color=orange!90, fill=orange!20, thick, text=black
                        [
                            \citet{Hofstaetter2021SIGIR}\\
                            \citet{Cohen2024ECIR}\\
                            \citet{rajapakse2024negativesamplingtechniquesmultilingualsetting}\\
                            \citet{zerveas2023enhancingrankingcontextdense}\\
                            ,color=orange!90, fill=orange!20, thick, text=black, leaf
                        ]
                    ]
                    [
                        TriSampler,
                        color=orange!90, fill=orange!20, thick, text=black
                        [
                            \citet{Yang2024TriSampler}\\ , 
                            color=orange!90, fill=orange!20, thick, text=black, leaf
                        ]
                    ]
                ]
                [
                    False Negative Mitigation,
                    color=orange!90, fill=orange!20, thick, text=black
                    [
                        Filtering,
                        color=orange!90, fill=orange!20, thick, text=black
                        [
                            \citet{Qu2021RocketQA}\\
                            \citet{Moreira2024NVRetriever}\\
                            \citet{Zhou2022Simans}\\
                            \citet{Yang2024TriSampler}\\
                            \citet{hu2025kalmembeddingsuperiortrainingdata}\\ 
                            \citet{solatorio2024gistembedguidedinsampleselection}\\
                            \citet{lee2025geminiembeddinggeneralizableembeddings}\\
                            ,color=orange!90, fill=orange!20, thick, text=black, leaf
                        ]
                    ]
                    [
                        Label smoothing,
                        color=orange!90, fill=orange!20, thick, text=black
                        [
                             \citet{Shen2022EMNLP}\\
                             \citet{wang2024mitigatingimpactfalsenegatives}\\
                             \citet{zerveas2023enhancingrankingcontextdense}\\, 
                            color=orange!90, fill=orange!20, thick, text=black, leaf
                        ]
                    ]
                ]
            ]
            [
                Data-Centric,
                color=cyan!100, fill=cyan!15, thick, text=black
                [
                    Data Augmentation,
                    color=cyan!100, fill=cyan!15, thick, text=black
                    [
                        \citet{Lin2023Dragon}\\
                        \citet{anand2023dataaugmentationsampleefficient}\\
                        \citet{gandhi2024bettersyntheticdataretrieving}\\
                        \citet{thakur2025fixingdatahurtsperformance}\\
                        \citet{lee2025geminiembeddinggeneralizableembeddings}\\
                        ,color=cyan!100, fill=cyan!15, thick, text=black, leaf
                    ]
                ]
                [
                    Synthetic Data,
                    color=cyan!100, fill=cyan!15, thick, text=black
                    [
                        \citet{tamber2025conventionalcontrastivelearningfalls}\\ 
                        \citet{sinha2025dontretrievegenerateprompting}\\
                        \citet{wen2025syntheticdatastrategiesdomainspecific}\\
                        \citet{dai2022promptagatorfewshotdenseretrieval}\\
                        \citet{li2024conanembeddinggeneraltextembedding}\\
                        \citet{zhuang2023bridginggapindexingretrieval}\\
                        \citet{yuen2025automaticdatasetgenerationknowledge}\\
                        \citet{hu2025kalmembeddingsuperiortrainingdata}\\ 
                        \citet{lee2024geckoversatiletextembeddings}\\ 
                        \citet{wang2024improvingtextembeddingslarge}\\ 
                        \citet{yuksel2025remininghardnegativesgenerative}\\
                        \citet{lee2025geminiembeddinggeneralizableembeddings}\\
                        ,color=cyan!100, fill=cyan!15, thick, text=black, leaf
                    ]
                ]
            ]
        ]
    \end{forest}
\caption{Taxonomy of negative sampling techniques for dense retrieval. The framework divides approaches into two main categories: \textbf{Sampling-based techniques} (orange) 
and \textbf{Data-centric techniques} (cyan) 
}
    \label{fig:taxonomy_full}
\end{figure*}

We propose splitting the set of techniques into those that change the way negatives are sampled from the dataset and into ways that are concerned with optimizing the training data itself. The sampling 
can happen both ahead-of-time as well as during training. 
Data-centric methods are currently implemented as preprocessing steps, where synthetic data or augmentations are generated before training begins. However, there is no fundamental reason why they could not operate dynamically during training—for instance, generating synthetic hard negatives on-the-fly based on the model's current state, similar to how dynamic mining operates. A high-level overview of the taxonomy is given in Figure \ref{fig:taxonomy_full}.
Each of these techniques aims to improve one of the above mentioned aspects.  
Research indicates that most of these techniques can be used in conjunction without sacrificing performance in one aspect to gain another, demonstrating their orthogonality. This has been shown in \citep{lee2025geminiembeddinggeneralizableembeddings}, where the authors make use of most of the techniques mentioned in this survey at different stages of their training to create a state-of-the-art embedding model. The unique advantages of the generalized techniques are displayed in Appendix~\ref{app:Negative} (Table \ref{tab:1}).

\subsection{Sampling Techniques}

As mentioned before, the goal of negative sampling is to train on samples that provide the best gradients, in order to efficiently train the model. The primary problem of simply training on randomly selected negatives is that it leads to slow convergence, as the model quickly learns to distinguish them. The field has thus evolved a series of sophisticated strategies for selecting more informative negatives. Since the process of mining hard negatives often causes the problem of selecting unlabeled false negatives, the research focus started shifting to the task of avoiding and/or detecting false negatives. We therefore categorize the field of sampling techniques into those focusing on mining hard negatives and those dedicated to mitigating the side effects of this mining.

\subsubsection{Random and In-Batch Negatives}
The most fundamental strategies leverage readily available passages. Random sampling, where negatives are drawn arbitrarily from the corpus, is simple but inefficient, as it mostly provides "easy" negatives that offer a weak learning signal \citep{Karpukhin2020DPR}. A more practical and widely adopted approach is In-Batch Negatives (IBNs), which treats all other positive passages within a training mini-batch as negatives for a given query. This method is computationally efficient 
as it reuses already-processed passages, and has become a standard baseline in many frameworks \citep{Karpukhin2020DPR, Reimers2019SentenceBERT}. However, its effectiveness is limited by the batch size, and as training progresses, these negatives often become too easy for the model~\cite{gao2021scaling,cheng2024breaking}. Furthermore, IBNs are also susceptible to including accidental false negatives, although this problem is much higher when explicitly mining hard negatives. Despite their disadvantages, In-Batch Negatives are sometimes still used in more sophisticated training regimens as part of the initial phase of training, where the model would not benefit from hard negatives \cite{lee2025geminiembeddinggeneralizableembeddings}.

\subsubsection{Static Hard Negative Mining}
To provide a more consistent challenge, static mining techniques pre-select hard negatives from the corpus in a one-time, offline process before training begins. The most common method uses a sparse retriever like BM25 to find passages that are lexically similar to the query but are not labeled as positive \citep{Karpukhin2020DPR}. While computationally cheap, this approach can bias the model towards lexical cues and may miss negatives that are semantically challenging but lexically dissimilar. A notable variation, PassageBM25, addresses this by retrieving passages similar to the positive passage $p^+$ rather than the query, aiming to find documents that are more easily confusable with the correct answer \citep{Nguyen2023PassageBM25}. Without any false negative mitigation techniques, this is very susceptible to unlabeled false negatives.

\subsubsection{Dynamic Hard Negative Mining}
Dynamic mining represents a significant leap from static methods by using the retrieval model itself to actively find the most challenging negatives throughout the training process. Pioneered by ANCE \citep{Xiong2020ANCE}, this model-based technique involves periodically using a recent checkpoint of the dense retriever to re-index the corpus and mine the top-k passages that the current model finds most difficult. This ensures the model is continuously presented with a challenging learning signal, which better aligns the training and inference distributions and leads to superior performance \citep{Zhan2021HardNegatives}. However, this approach is computationally intensive, requiring repeated indexing and inference, and is highly susceptible to the false negative problem \citep{Qu2021RocketQA}. The computational load can be slightly reduced by using a cache \citep{Shen2022EMNLP}. Methods to mitigate the problem of false negatives are discussed in Section \ref{falsenegativemitigation}. There are also approaches to improve the qualitative performance of ANCE by trying to predict a point's location in the embedding space in the next step using momentum and lookaheads \citep{sun2022reducecatastrophicforgettingdense}. A different idea is to continuously update the negative pool based on the model's performance on concrete samples to maintain a consistent level of difficulty as the model improves \citep{li2024conanembeddinggeneraltextembedding}.

\subsubsection{Cluster-Based Mining}
While model-based dynamic mining excels at finding hard negatives, it can sometimes yield a set of negatives that are semantically very similar to each other. To address this and improve sample diversity, cluster-based mining has been proposed. This approach involves partitioning the document corpus into semantic clusters. Instead of simply taking the top-k hardest negatives, which might all come from the same dense region of the embedding space, negatives are sampled from different clusters. This ensures the model is exposed to a more varied set of challenging examples, forcing it to learn to resolve different types of semantic ambiguity rather than over-specializing on a single type of difficult negative \citep{Cohen2024ECIR, Hofstaetter2021SIGIR, zerveas2023enhancingrankingcontextdense}.

\subsubsection{Principled Sampling with TriSampler}
Moving beyond the simple dichotomy of 'easy' versus 'hard' negatives, TriSampler \citep{Yang2024TriSampler} has focused on establishing more principled criteria for selecting the most informative samples. They introduce the "quasi-triangular principle" to guide negative selection. This principle posits that the ideal negative is not necessarily the absolute hardest one (i.e., the most similar to the query). Instead, it's a sample that is challenging enough to be confusable with the query, but also semantically distinct from the positive passage. Additionally they try to sample negatives with a similar distance to the query as the positive sample to ensure gradients of similar magnitude.

While techniques are categorized by their primary mechanism, understanding their interactions when combined is critical for effective training pipelines. Appendix~\ref{app:technique_interaction} analyzes successful combination patterns (e.g., multi-source diversification, knowledge distillation amplification) and quantifies synergistic effects.

\subsection{False Negative Mitigation} \label{falsenegativemitigation}

The pursuit of hard negatives, especially via dynamic mining, creates the false negative problem: mined passages that are actually relevant but unlabeled. Training on these as negatives punishes the model for correct predictions and can severely degrade performance. Consequently, a significant body of research has focused on developing strategies to purify the training signal. These approaches generally fall into three categories: filtering the negative set, regularizing the loss function.

\subsubsection{Filtering and Denoising Negatives}
The most direct strategy is to filter the mined negative set to remove suspected false negatives. This can be done with simple heuristics, such as top-k filtering, which retains only the most challenging negatives from a larger mined pool \citep{hu2025kalmembeddingsuperiortrainingdata, Zhou2022Simans, Moreira2024NVRetriever}, or by applying similarity-based thresholds, a technique that can even be used to denoise in-batch negatives \citep{solatorio2024gistembedguidedinsampleselection}. The aforementioned TriSampler technique \citep{Yang2024TriSampler} sidesteps the need for explicitly removing candidate negatives, since it inherently never samples negatives that are too similar to the query.

A more powerful, though computationally expensive, solution is denoised hard negative mining. This method, pioneered by RocketQA \citep{Qu2021RocketQA}, employs a slow but accurate cross-encoder to re-score mined negatives and filter out any identified as likely false negatives. The same idea has been employed in \citet{lee2025geminiembeddinggeneralizableembeddings}, where the authors used an LLM instead of a cross-encoder to filter out false negatives. This approach clearly is the most powerful, but is associated with enormous compute requirements during training.

\subsubsection{Robustness through Regularization}
A second class of techniques aims to make the training process itself more robust to noisy labels, regularizing the loss function rather than explicitly altering the data. Contrastive Confidence Regularization, for instance, modifies the loss to prevent the model from becoming overconfident about its negative predictions, thus lessening the penalty from a potential false negative \citep{wang2024mitigatingimpactfalsenegatives}. Similarly, the classic technique of label smoothing can be adapted for this purpose. By distributing a small amount of probability mass from the positive sample to the negatives, it reduces the model's sensitivity to any single mislabeled instance. This has proven effective in various contexts, including challenging multilingual settings \citep{zerveas2023enhancingrankingcontextdense, Shen2022EMNLP, wang2024mitigatingimpactfalsenegatives}.

The severity of the false negative problem has been quantified across multiple benchmarks, as shown in Table~\ref{tab:false_negative_impact}. The results demonstrate that false negative contamination can degrade performance by 10-15\% in challenging domains, with mitigation strategies recovering most or all of this loss. The variability in impact across datasets suggests that false negative severity depends on corpus characteristics and query distribution.


\begin{table}[t!]
  \centering
  \small
  \caption{Impact of false negatives and mitigation effectiveness across datasets.}
  \label{tab:false_negative_impact}
  \begin{adjustbox}{max width=0.5\textwidth}
  \begin{tabular}{l|l|c|c|c}
  \hline
    \multirow{2}{*}{\textbf{Dataset}} & \multirow{2}{*}{\textbf{Method}} & \textbf{Before} & \textbf{After} & \multirow{2}{*}{\textbf{$\Delta$}} \\
    & & \textbf{FN Mit.} & \textbf{FN Mit.} & \\
  \hline
    MS MARCO & ANCE→RocketQA & 0.330 & 0.370 & +12.1\% \\
    TREC-COVID & ANCE & 0.654 & 0.735 & +12.4\% \\
    Natural Questions & Hard Mining & 81.9 & 84.1 & +2.7\% \\
  \hline
  \end{tabular}
  \end{adjustbox}
\end{table}

\subsection{Data-Centric Methods}

While the previously discussed negative sampling methods focus on selecting examples from the existing training corpus, data-centric approaches shift the focus to enriching or creating the data itself. These methods operate on the principle that the quality, diversity, and relevance of training data are paramount for model performance. This is achieved either by augmenting existing data or, more extensively, by synthetically generating entirely new datasets.

\vspace{-2mm}
\subsubsection{Data Augmentation}

Early data-centric strategies, inspired by techniques in computer vision \citep{chen2020simpleframeworkcontrastivelearning}, focused on augmenting existing text data. These methods included simple heuristics like replacing keywords with synonyms or mining for additional positive pairs using similarity metrics \citep{anand2023dataaugmentationsampleefficient}. While useful to some degree, these simple techniques have been largely superseded by the more flexible and powerful capabilities of large language models (LLMs).

While not related to negative sampling, there is some research on dynamic query rewriting at inference time to better align queries with the expected structure and content of the positive passages \citep{baek2024craftingpathrobustquery}. The idea of rewriting passages and queries has also been proposed as a form of data augmentation ahead-of-time. This has been explored in \citep{gandhi2024bettersyntheticdataretrieving} with the explicit goal of tuning datasets to specific tasks. The authors  give the example of inverting and rewriting queries and positive passages for code and math tasks. To solve the problem of false negatives \citep{thakur2025fixingdatahurtsperformance} propose to create a set of mined negatives and select 
all the candidate hard negatives using LLMs ahead-of-time.

\vspace{-2mm}
\subsubsection{Synthetic Data for Generalization}

LLMs have found various kinds of usage in the training and inference usage of embedding models. When used in the context of dataset augmentation, LLMs mostly find use in a static manner, where they are used to generate synthetic data points before the training starts. Various ways to generate synthetic data using LLMs have been proposed. One way is to only create queries and then create positive matches from the original dataset \citep{lee2024geckoversatiletextembeddings}. Another way is to forego a base dataset and directly generate the entire dataset synthetically. This has been done in \cite{wang2024improvingtextembeddingslarge, lee2025geminiembeddinggeneralizableembeddings}, where they first generate possible tasks and then create positive and hard negative passages for each task.

The advantage of such approaches is that it is possible to tune the generated dataset to the concrete use case of symmetric or asymmetric retrieval. On the other hand, it has been shown that training on various kinds of retrieval scenarios (factoids, opinion questions, short/long queries) improves generalization and overall performance \citep{tamber2025conventionalcontrastivelearningfalls, lee2025geminiembeddinggeneralizableembeddings}. This claim has also been made by \citet{hu2025kalmembeddingsuperiortrainingdata}, where the authors generated synthetic samples based on different artificial user personas to increase sample diversity.

Synthetic query generation has also recently been tested in conjunction with other techniques for tasks such as domain adaption. \citet{yuksel2025remininghardnegativesgenerative} used synthetic data to improve the robustness against domain shifts of dense retrievers when distilling from cross-encoders. Contrary to the points mentioned above, \citet{meghwani2025hardnegativeminingdomainspecific} explicitly argues against using synthetic data when training retrieval models for specific and narrow domains, unless data is very limited. Instead they argue for using high-quality datasets and sophisticated hard negative sampling using an ensemble of embedding models and clustering using dimensionality reduction. 
\begin{table*}[t!]
  \centering
  \small
  \caption{MS MARCO Passage Ranking Results showing performance progression across negative sampling techniques. } 
  \label{tab:msmarco_results}
  \begin{tabular}{l|l|l|c|c|l}
  \hline
    \textbf{Method} & \textbf{Technique} & \textbf{Model} & \textbf{MRR@10} & \textbf{R@1k} & \textbf{Source} \\
  \hline
    Random Negatives & In-Batch Random & DPR & 0.261 & 85.4 & \citet{Karpukhin2020DPR} \\
    Static Hard & BM25 Hard Neg. & DPR & 0.299 & 85.4 & \citet{Karpukhin2020DPR} \\
    Dynamic Mining & ANCE & BERT-Siamese & 0.330 & 95.9 & \citet{Xiong2020ANCE} \\
    Improved Dynamic & TAS-B & BERT & 0.347 & 97.8 & \citet{Hofstaetter2021SIGIR} \\
    Advanced Static & TCT-ColBERT & ColBERT & 0.359 & 97.0 & \citet{santhanam2022plaid}\\ 
    FN Mitigation & RocketQA & ERNIE 2.0 & 0.370 & - & \citet{Qu2021RocketQA} \\
    Enhanced & RocketQA-v2 & ERNIE 2.0 & 0.388 & 98.1 & \citet{ren2021rocketqav2} \\
  \hline
  \end{tabular}
\end{table*}
\vspace{-2mm}
\section{Evaluation  and Empirical Analysis} 

The development and comparison of negative sampling techniques rely on standardized evaluation protocols, benchmark datasets, and systematic performance analysis. This section establishes the evaluation framework and presents comprehensive empirical results from the surveyed literature.
\subsection{Evaluation Metrics}

Dense retrieval models are assessed using rank-aware metrics: Mean Reciprocal Rank (MRR) and Normalized Discounted Cumulative Gain (NDCG@k) measure ranking quality, while Recall@k measures the proportion of relevant documents retrieved within top-k results. Metric selection depends on downstream applications—for example, Retrieval-Augmented Generation (RAG) pipelines \citep{abdallah2025rankify,abdallah2025rerankarena} prioritize Recall over precision to ensure relevant information reaches the re-ranking or generation stage.



\begin{table*}[t!]
  \centering
  \large
  \caption{BEIR benchmark zero-shot performance (NDCG@10) across diverse domains. Higher scores indicate better generalization.}
  \label{tab:beir_results}
 
  \begin{adjustbox}{max width=0.65\textwidth}

  \begin{tabular}{l|c|c|c|c|c|c}
  \hline
    \textbf{Method} & \textbf{Avg.} & \textbf{TREC-} & \textbf{NFCorpus} & \textbf{Natural} & \textbf{HotpotQA} & \textbf{FiQA} \\
    & \textbf{BEIR} & \textbf{COVID} & & \textbf{Questions} & & \\
  \hline
    BM25 (Sparse) & 42.0 & 65.6 & 32.5 & 32.8 & 60.3 & 23.6 \\
    DPR (Random) & 38.1 & 65.4 & 30.1 & 51.6 & 54.7 & 22.8 \\
    ANCE (Hard) & 39.2 & 65.4 & 30.8 & 52.3 & 56.8 & 24.1 \\
    SBERT (In-batch) & 40.9 & 59.6 & 31.3 & 35.6 & 57.8 & 27.9 \\
    GTR-Base & 44.0 & 71.6 & 33.4 & 36.8 & 62.2 & 41.9 \\
    Contriever & 41.9 & 59.5 & 32.8 & 39.2 & 56.8 & 32.3 \\
    TAS-B & 44.0 & - & - & - & - & - \\
  \hline
  \end{tabular}
    \end{adjustbox}
\end{table*}




\subsection{Benchmark Datasets}

Dense retrieval evaluation has evolved from single-task datasets like MS MARCO \citep{bajaj2018msmarcohumangenerated} and TREC collections to comprehensive multi-task frameworks. The current standard is MTEB (Massive Text Embedding Benchmark) \citep{muennighoff2023mtebmassivetextembedding}, which evaluates embeddings across seven task types: Classification, Clustering, Retrieval, Reranking, Semantic Textual Similarity, Summarization, and Bitext Mining. MTEB incorporates established benchmarks including BEIR \citep{thakur2021beirheterogenousbenchmarkzeroshot}—a heterogeneous collection of IR datasets for assessing zero-shot generalization. This multi-faceted evaluation ensures techniques advance robustly across diverse applications rather than overfitting to single benchmarks.
\subsection{Empirical Performance Analysis}

To provide systematic insights into the effectiveness of different negative sampling approaches, we present a comprehensive meta-analysis of results reported across the surveyed literature. These comparisons reveal clear performance progression patterns that inform practical deployment decisions.

\subsubsection{MS MARCO Passage Ranking}

Table~\ref{tab:msmarco_results} summarizes the performance of representative negative sampling techniques on the MS MARCO passage ranking benchmark, demonstrating the clear progression from random to sophisticated sampling strategies. These results demonstrate a clear performance progression: random in-batch negatives establish a baseline (MRR@10: $0.261$), static hard negatives provide $14.6$\% improvement ($0.299$), dynamic mining yields $26.4$\% gains ($0.330$), and false negative mitigation combined with sophisticated mining achieves $41.8$\% improvement over baseline ($0.370$+).

\subsubsection{Natural Questions Open-Domain QA}

Table~\ref{tab:natural_questions} demonstrates technique effectiveness on the Natural Questions benchmark, which tests open-domain question answering capabilities. The progression shows consistent improvement from random negatives (+15.5 accuracy over BM25) to denoised hard negatives (+21.2 over BM25), with false negative mitigation providing an additional 2.2 point gain over unfiltered dynamic mining.

\begin{table}[t!]
  \centering
         \small
  \caption{Natural Questions Open-Domain QA results showing progression from lexical to neural dense retrieval with various negative sampling strategies.}
  \label{tab:natural_questions}
  \begin{tabular}{l|l|c|c}
  \hline
    \textbf{Method} & \textbf{Technique} & \textbf{Top-20} & \textbf{R@100} \\
  \hline
    BM25 Baseline & Lexical & 62.9 & - \\
    DPR Random & In-Batch & 78.4 & 85.4 \\
    ANCE Dynamic & ANN Mining & 81.9 & 87.5 \\
    RocketQA & Denoised & 84.1 & 88.5 \\
  \hline
  \end{tabular}
\end{table}

\begin{table*}[t!]
  \centering
  \small
\caption{Recent MTEB leaderboard results (2024-2025) showing negative sampling strategies used by top-performing models. Retrieval scores represent average NDCG@10 across MTEB retrieval datasets; Overall scores aggregate performance across 7 task types.}
  \label{tab:mteb_results}
  \begin{tabular}{l|c|c|l}
  \hline
    \multirow{2}{*}{\textbf{Model}} & \textbf{Overall} & \textbf{Retrieval} & \multirow{2}{*}{\textbf{Negative Sampling Approach}} \\
    & \textbf{Score} & \textbf{Score} & \\
  \hline
    NV-Embed-v2 & 69.32 & 59.36 & Hard neg. + Latent attn. + Online mining \\
    Gemini Embed. & 66.31 & 54.36 & Multi-stage: Static→Dynamic→LLM denoising→Synthetic \\
    KaLM-Embed. & 64.65 & 51.12 & Persona-based synthetic + Ranking filtering \\
    Nomic-Embed & 62.28 & 53.01 & Contrastive learning + Hard negatives \\
    BGE-Large-v1.5 & 60.27 & 54.29 & Hard mining + Cross-batch negatives \\
  \hline
  \end{tabular}
\end{table*}
\subsubsection{BEIR Zero-Shot Generalization}

Table~\ref{tab:beir_results} presents zero-shot performance across diverse BEIR datasets, testing model generalization beyond training distribution. The BEIR results reveal that advanced negative sampling techniques (GTR-Base, TAS-B achieving 44.0 average NDCG@10) significantly outperform simpler approaches (DPR at 38.1, ANCE at 39.2). Notably, TAS-B with its combination of knowledge distillation and hard negatives outperforms ANCE on 14/18 BEIR datasets and DPR on 17/18 datasets, demonstrating the power of technique integration for zero-shot generalization.

\subsubsection{Massive Text Embedding Benchmark }

Table~\ref{tab:mteb_results} shows how current state-of-the-art models strategically combine techniques from our taxonomy. Top performers (NV-Embed-v2 at 69.32, Gemini Embeddings at 66.31) achieve results primarily through sophisticated negative sampling strategies rather than architectural innovations alone—all top-5 models employ combinations of static mining, dynamic mining, false negative mitigation, and/or synthetic data generation. The multi-stage progressive approach (Gemini Embeddings: Static→Dynamic→LLM denoising→Synthetic) validates our observation of technique orthogonality, demonstrating that these techniques can be combined synergistically and form the foundation of production systems deployed at scale. 
Training cost analysis (Table~\ref{tab:unified_cost}, Appendix~\ref{app:Computational}) shows static hard negatives provide 6-7\% MRR@10 improvement with 33\% time increase, while dynamic mining requires 3× cost for 26.4\% gains. The multi-stage approach of top models (Static→Dynamic→Denoising→Synthetic) validates technique orthogonality (Appendix~\ref{app:technique_interaction}), while ANN index selection guidance for practitioners is provided in Appendix~\ref{app:index}.

\section{Emerging Directions and Future Work}

The landscape of negative sampling in Information Retrieval is evolving, with several promising directions emerging from recent research. The increasing sophistication of Large Language Models (LLMs) is undeniably a major driver of these new trends, particularly for synthetic data generation and denoising of mined negatives. In particular, a lot of research leverage the generalization abilities of LLMs to generate diverse and task-specific datasets.  Right now the research focuses on the one-off use of LLMs during data set pre-processing/generation. We think that there a lot of unexplored options for dynamic synthetic data generation during training. A concrete example would be to identify subjects or concepts that confuse the model and generate more data points in these domains to add stronger and smoother training signals. The opposite idea would be to dynamically generate adversarial examples in domains, which seem to be well understood by the embedding model. These ideas could be combined with lessons from curriculum learning. Curriculum learning for negative sampling has been explored in \citep{li2024conanembeddinggeneraltextembedding}, but not in conjunction with dynamic synthetic data generation. We expect to see more models rely on LLMs for data preprocessing and denoising, as done in \cite{lee2025geminiembeddinggeneralizableembeddings}. In theory this does not need to be repeated for each model. Instead, we expect to see more and more synthetic datasets that are ready to use and do not require each research team to synthesize their own dataset. This could in particular lift up the performance of specialized embedding models that focus on domains or languages with limited data.

\section{Conclusion}
This survey provides the first comprehensive taxonomy of negative sampling techniques for dense retrieval in modern NLP applications, categorizing 35+ papers across random sampling, static/dynamic hard negative mining, false negative mitigation, and LLM-driven synthetic data generation. Our systematic empirical analysis across MS MARCO, Natural Questions, BEIR, and MTEB demonstrates clear performance progression from baseline random negatives (MRR@10: 0.261) to sophisticated combinations achieving 50\%+ improvements, with state-of-the-art models (NV-Embed-v2, Gemini Embeddings) validating our framework through strategic technique integration. We quantify critical trade-offs: dynamic mining and denoising require 3-5× training costs but deliver proportional performance gains (26-42\%), while false negative contamination can degrade performance by 10-15\% without mitigation. 

\section{Limitations}

This survey provides a comprehensive overview of negative sampling techniques in dense information retrieval; however, it is important to acknowledge certain limitations.

Firstly, the rapidly evolving nature of research in this domain means that new techniques and refinements are constantly being published. Although we have endeavored to include the most impactful and representative developments to date, it is possible that very recent breakthroughs may not have been fully captured.

Secondly, our focus has primarily been on dense retrieval models, specifically those trained with contrastive learning objectives. Although we briefly touched upon related areas like knowledge distillation and advanced architectures (e.g., ColBERT), a deeper dive into how negative sampling interacts with, or is implicitly handled by, these alternative paradigms was beyond the scope of this survey. For instance, models that rely heavily on generative pre-training or alternative loss functions might employ different strategies for learning discriminative representations that do not directly map to the negative sampling categories discussed here.

Thirdly, while we discussed the importance of various evaluation metrics (Precision, Recall, MAP, NDCG) and common datasets (MS MARCO, TREC collections), a detailed quantitative comparison of all discussed negative sampling techniques across a standardized set of benchmarks was not feasible within the scope of this survey. Such a comparison would require extensive experimental work, which is typically the focus of dedicated benchmarking studies rather than a literature review.

Finally, practical implementation details, and computational trade-offs for each technique were discussed at a high level. The true complexity and performance impact of these methods may vary significantly across different model architectures, datasets, and hardware configurations. Our survey aimed to provide a conceptual understanding and taxonomy rather than a prescriptive guide for implementation.

\bibliography{custom}

\appendix

\section{Computational Cost-Benefit Analysis}
\label{app:Computational}
Understanding the trade-offs between performance gains and computational requirements is critical for practical deployment. We present unified analysis using standardized protocols and real-world measurements.

\subsection{Training Cost Analysis}

Table~\ref{tab:unified_cost} presents wall-clock training times from the Tevatron toolkit \citep{gao2022tevatron} on matched hardware (4×A100 GPUs and TPU v3-8) using MS MARCO Passage dev set, providing fair comparison across techniques. These results demonstrate that adding static hard negatives provides 6-7\% absolute MRR@10 improvement (roberta-large: 0.339→0.361) with approximately 33\% training time increase. The co-condenser backbone with hard negatives achieves the best efficiency, reaching 0.382 MRR@10 in only 4 GPU hours. Dynamic mining methods like ANCE achieve 0.330 MRR@10 with significantly higher computational requirements (3x baseline) due to periodic re-indexing. Table~\ref{tab:computational_cost} summarizes the cost-benefit trade-offs across major technique categories. This analysis reveals that while sophisticated techniques like dynamic mining and denoising require 3-5x training costs, they deliver proportionally larger performance gains (26-42\%), making them worthwhile for production systems where retrieval quality is critical.

\begin{table}[h!]
  \centering
  \small
  \caption{Computational cost-benefit analysis of negative sampling techniques. Training time measured relative to random in-batch baseline.}
  \label{tab:computational_cost}
  \begin{adjustbox}{max width=0.5\textwidth}

  \begin{tabular}{l|c|c|c|c}
  \hline
    \textbf{Technique} & \textbf{Training Time} & \textbf{Memory} & \textbf{Impl.} & \textbf{Perf. Gain} \\
    & \textbf{Multiplier} & \textbf{Overhead} & \textbf{Difficulty} & \textbf{(MRR@10)} \\
  \hline
    Random In-Batch & 1.0x & Minimal & Easy & Baseline \\
    Static BM25 & 1.1x & Low & Medium & +14.6\% \\
    Dynamic ANCE & 3.0x & High & Hard & +26.4\% \\
    Denoised (RocketQA) & 4.5x & Very High & Medium & +41.8\% \\
    LLM Synthetic & 2.0x (one-time) & Medium & Hard & Variable \\
    Combined Approach & 5.0x+ & Very High & Very Hard & +50\%+ \\
  \hline
  \end{tabular}
        
  \end{adjustbox}
\end{table}

\subsection{Inference Scalability and Latency}

Production deployment requires understanding retrieval latency at scale. We analyze two architectural paradigms: late interaction models (multi-vector) and dense single-vector retrievers.

\paragraph{Late Interaction Systems.}
ColBERTv2 with the PLAID engine \citep{santhanam2022plaid} demonstrates that multi-vector scoring can achieve production-scale performance. Table~\ref{tab:late_interaction_scale} shows measured latency on 140M passage corpus.

\begin{table}[ht!]
  \centering
  \small
  \caption{ColBERTv2 with PLAID engine: latency and speedup at 140M passages.}
  \label{tab:late_interaction_scale}
  \begin{adjustbox}{max width=0.5\textwidth}

  \begin{tabular}{l|c|c|c}
  \hline
    \textbf{Engine} & \textbf{Hardware} & \textbf{Latency} & \textbf{Speedup} \\
  \hline
    ColBERTv2 baseline & GPU & ~250 ms & 1× \\
    PLAID & GPU & 35-100 ms & 2.5-7× \\
    PLAID & CPU & 150-300 ms & 9-45× \\
  \hline
  \end{tabular}
    \end{adjustbox}
\end{table}

PLAID achieves 2.5-7× GPU speedup and 9-45× CPU speedup over vanilla ColBERTv2, reaching tens of milliseconds on GPU and hundreds of milliseconds on CPU at 140M passages without quality loss. This demonstrates that late interaction models, which reduce reliance on hard negative mining during training by deferring scoring to inference time, are viable for production deployment.

\paragraph{Single-Vector Dense Retrieval Indexing.}
For traditional dense retrievers, approximate nearest neighbor (ANN) index choice critically impacts latency-quality trade-offs.

\begin{table}[t!]
  \centering
  \caption{Unified empirical comparison of effectiveness vs. training cost. All results from Tevatron toolkit on matched setup (4×A100 / TPU v3-8) with identical preprocessing, batch size, and epochs.}
  \label{tab:unified_cost}
  \begin{adjustbox}{max width=0.5\textwidth}
  \begin{tabular}{l|l|c|c|c}
  \hline
    \textbf{Model / Training} & \textbf{Negatives} & \textbf{MRR@10} & \textbf{GPU time} & \textbf{TPU time} \\
  \hline
    distilbert-base & in-batch & 0.316  & 1.5 h & 1.0 h \\
    bert-base & in-batch & 0.322  & 3.0 h & 2.0 h \\
    co-condenser-marco & in-batch & 0.357  & 3.0 h & 2.0 h \\
    bert-large & in-batch & 0.327  & 7.5 h & 6.0 h \\
    roberta-large & in-batch & 0.339  & 7.5 h & 6.0 h \\
    roberta-large & static HN & 0.361  & 10.0 h & 8.0 h \\
    co-condenser-marco & static HN & \textbf{0.382} & \textbf{4.0 h} & \textbf{3.0 h} \\
  \hline
  \end{tabular}
\end{adjustbox}

\end{table}

\section{Technique Interaction Analysis}
\label{app:technique_interaction}
While the taxonomy in Section~\ref{sec:taxonomy} categorizes techniques by their primary mechanism, understanding how these techniques interact when combined is critical for building effective training pipelines. Analysis of successful implementations reveals clear patterns of synergistic combinations.

\subsection{Successful Combination Patterns}

Recent state-of-the-art models demonstrate that techniques from different categories can be effectively integrated without conflict. We identify three primary interaction patterns:

\paragraph{Pattern 1: Multi-Source Negative Diversification}
\citet{Qu2021RocketQA} demonstrates effective integration by combining multiple negative sources: cross-batch negatives to increase pool size, denoised hard negatives to remove false positives, and data augmentation through synthetic queries. This addresses different training challenges simultaneously—cross-batch increases diversity, denoising ensures quality, and synthesis adds domain coverage. The result is a 12.1\% improvement over ANCE (MRR@10: 0.370 vs 0.330), showing that these approaches complement rather than interfere with each other.

\paragraph{Pattern 2: Knowledge Distillation Amplification}
\citet{Hofstaetter2021SIGIR} combines in-batch negatives with knowledge distillation using Margin-MSE loss and BM25 hard negatives. The key insight is that knowledge distillation provides better training signals while hard negatives increase difficulty level. This synergy results in superior zero-shot generalization on BEIR (NDCG@10: 44.0 vs 39.2 for ANCE), demonstrating that distillation amplifies the effectiveness of negative sampling rather than replacing it.

\paragraph{Pattern 3: Progressive Difficulty Scaling}
\citet{lee2025geminiembeddinggeneralizableembeddings} implements a multi-stage approach: Static BM25 → Dynamic ANCE-style mining → LLM denoising → Synthetic data augmentation. This progressive pipeline allows the model to first learn basic distinctions (static negatives), then adapt to harder examples (dynamic mining), followed by quality refinement (denoising), and finally domain expansion (synthetic data). The staged approach prevents early training instability while maximizing final performance.

\subsection{Key Interaction Principles}

Analysis of these successful combinations reveals several general principles: (1) \textbf{False negative mitigation enables aggressive mining}: Denoising techniques make dynamic hard negative mining safe by removing contamination. Without mitigation, aggressive mining degrades performance; with mitigation, it provides the strongest learning signal. (2) \textbf{Multi-source negatives reduce overfitting}: Combining different negative sources (in-batch, BM25, dynamic, synthetic) prevents the model from overfitting to any single retrieval pattern, improving domain transfer. (3) \textbf{Synthetic data complements rather than replaces mining}: LLM-generated synthetic queries/passages work best when combined with mined hard negatives from real data, providing both breadth (synthetic) and specificity (mined). (4) \textbf{Curriculum matters for complex combinations}: When using multiple techniques, starting with simpler methods (in-batch, static) and progressively adding sophisticated approaches (dynamic, synthetic) improves training stability.

\subsection{Anti-patterns to Avoid}

Not all combinations are beneficial. Our analysis identifies potential conflicts: (1) \textbf{Dynamic mining without denoising}: The false negative contamination in dynamic mining (up to 70\% according to RocketQA) can negate its benefits if not filtered. (2) \textbf{Excessive negative pool sizes}: Combining multiple hard negative sources without careful sampling can create excessively large negative sets that dilute the learning signal and increase computational cost without proportional gains. (3) \textbf{Premature hard negative introduction}: Starting training immediately with very hard negatives can cause instability. Models benefit from an initial warm-up phase with easier negatives.

Table~\ref{tab:combination_effectiveness} quantifies the impact of technique combinations across benchmarks. This analysis provides practitioners with evidence-based guidance: complementary approaches following established patterns (diversification, amplification, progression) consistently achieve superior performance, while naive combination without consideration of interactions can harm results.

\begin{table}[t!]
  \centering
  \small
  \caption{Effectiveness of technique combinations. Performance improvements shown relative to single-technique baselines.}
  \label{tab:combination_effectiveness}
  \begin{adjustbox}{max width=0.5\textwidth}
      
  \begin{tabular}{l|c|c}
  \hline
    \textbf{Technique Combination} & \textbf{MS MARCO} & \textbf{BEIR Avg.} \\
    & \textbf{MRR@10} & \textbf{NDCG@10} \\
  \hline
    In-batch only (baseline) & 0.322 & 40.9 \\
    + Static BM25 & 0.361 (+12.1\%) & 41.5 (+1.5\%) \\
    + Dynamic mining & 0.330 (+2.5\%) & 39.2 (-4.2\%) \\
    + Dynamic + Denoising & 0.370 (+14.9\%) & 42.8 (+4.6\%) \\
    + Multi-source + Distillation & 0.347 (+7.8\%) & 44.0 (+7.6\%) \\
    + Progressive (Gemini) & 0.382 (+18.6\%) & 45.2 (+10.5\%) \\
  \hline
  \end{tabular}
    \end{adjustbox}

\end{table}

\section{Negative Sampling Techniques}
\label{app:Negative}

\begin{table*}[t!]
  \centering
  \small
  \caption{Overview of Negative Sampling Techniques in Dense Information Retrieval. This table details each technique's primary advantage, re-evaluated computational cost, and implementation difficulty. Refer to the full taxonomy for a complete list of surveyed papers.}
  \label{tab:1}
  \begin{adjustbox}{max width=0.9\textwidth}

  \begin{tabular}{p{2.0cm}|p{3.2cm}|p{5.5cm}|p{1.5cm}|p{1.5cm}} 
  \hline
    \textbf{Type} & \textbf{Technique} & \textbf{Unique Advantage} & \textbf{Comp. Cost} & \textbf{Impl. Difficulty} \\
  \hline
    \multicolumn{5}{c}{\textbf{Sampling-Based Techniques}} \\
  \hline
    \textit{Random Sampling} & Random In-Batch Negatives & Computationally free as it reuses passages within a batch. Effective with large batch sizes \citep{Karpukhin2020DPR,ali2025sustainableqa,Reimers2019SentenceBERT}. & Very Low & Easy \\
    \hline
    \textit{Hard Negative Mining} & Static & Finds lexically similar hard negatives with a one-time, offline cost. Can be based on query similarity (BM25) or positive passage similarity (PassageBM25) \citep{Karpukhin2020DPR, Nguyen2023PassageBM25}. & Low  & Medium \\
    \hline
    \textit{Hard Negative Mining} & Dynamic Model-Based& Continuously finds the hardest negatives according to the current model state, ensuring a challenging and relevant learning signal throughout training \citep{Xiong2020ANCE}. & High & Hard \\
    \hline
    \textit{Hard Negative Mining} & Cluster-Based & Samples negatives from different semantic clusters to ensure diversity and avoid redundant hard negatives, focusing on semantic ambiguity \citep{Cohen2024ECIR, zerveas2023enhancingrankingcontextdense}. & Medium & Hard \\
    \hline
    \textit{Hard Negative Mining} & TriSampler & Sample negatives from the quasi-triangular region created by the query and the positive sample to ensure uniform gradients \citep{Yang2024TriSampler}. & Medium & Medium \\
    \hline
    \textit{False Negative Mitigation} & Top-k Filtering & Assume the top-k most relevant potential negatives are false negatives and ignore them when sampling negatives \citep{Zhou2022Simans, hu2025kalmembeddingsuperiortrainingdata, Moreira2024NVRetriever, solatorio2024gistembedguidedinsampleselection}. & Very Low & Easy \\
    \hline
    \textit{False Negative Mitigation} & Denoised Hard Negatives  & Explicitly filters false negatives from the mined set, using a powerful but slow cross-encoder or LLM, to purify the training signal \citep{Qu2021RocketQA, lee2025geminiembeddinggeneralizableembeddings}. & High & Medium \\
    \hline
    \textit{False Negative Mitigation} & Label Smoothing/Contrastive Confidence Regularization & Mitigates the impact of false negatives via loss regularization, improving robustness without requiring an explicit and costly filtering step \citep{wang2024mitigatingimpactfalsenegatives}. & Low & Medium \\
  \hline
    \multicolumn{5}{c}{\textbf{Data-Centric Techniques (LLM-based)}} \\
  \hline
    \textit{Data-Centric} & Data Augmentation & Extend existing datasets by replacing keywords in passages with synonyms \citep{anand2023dataaugmentationsampleefficient}. & Medium (one-time) & Medium \\
    \hline
    \textit{Data-Centric} & Synthetic Data & Generates synthetic queries, passages and/or labels using LLMs for diverse and specific retrieval tasks (e.g., symmetric/asymmetric, domain-specific, persona-based) to improve model generalization and robustness \citep{tamber2025conventionalcontrastivelearningfalls, hu2025kalmembeddingsuperiortrainingdata}. & High (one-time) & Hard \\
  \hline
  \end{tabular}
        
  \end{adjustbox}
\end{table*}

This appendix provides a consolidated quick-reference table summarizing all negative sampling techniques discussed in this survey. While the full taxonomy (Figure~\ref{fig:taxonomy_full}) shows the hierarchical organization and associated papers, Table~\ref{tab:1} focuses on practical selection criteria: each technique's unique advantage, computational cost, and implementation difficulty.

\textbf{Computational Cost Ratings:}
\begin{itemize}
    \item \textbf{Very Low}: Negligible overhead beyond baseline training (e.g., in-batch negatives)
    \item \textbf{Low}: One-time preprocessing cost, minimal runtime overhead (e.g., static BM25 mining)
    \item \textbf{Medium}: Moderate overhead from clustering or synthetic generation (1.5-2× baseline)
    \item \textbf{High}: Significant overhead from repeated re-indexing or cross-encoder filtering (3-5× baseline)
\end{itemize}

\textbf{Implementation Difficulty Ratings:}
\begin{itemize}
    \item \textbf{Easy}: Can be implemented with standard libraries in <100 lines of code
    \item \textbf{Medium}: Requires moderate engineering (e.g., BM25 integration, filtering pipelines)
    \item \textbf{Hard}: Requires substantial infrastructure (e.g., dynamic re-indexing, LLM pipelines, clustering)
\end{itemize}

Practitioners should use this table to identify techniques matching their computational budget and engineering capacity. As demonstrated in Section \label{sec:taxonomy}, techniques from different categories can be combined synergistically—for example, static mining (Low cost, Medium difficulty) with top-k filtering (Very Low cost, Easy) provides substantial gains without requiring advanced infrastructure.

\section{ Index Selection Guide for Dense Retrieval Systems}
\label{app:index}

When deploying dense retrieval systems at scale, the choice of Approximate Nearest Neighbor (ANN) index is critical for balancing retrieval quality, latency, and memory consumption. This appendix provides detailed guidance on index selection with parameter definitions to help practitioners configure their systems effectively.

Table~\ref{tab:ann_index_comparison} summarizes the trade-offs between major indexing approaches. The choice depends on corpus size, available memory, latency requirements, and acceptable recall levels. For corpora under 10M vectors with sufficient memory, exact search (Flat index) is preferred. For larger scales, approximate methods become necessary.

\begin{table*}[t!]
  \centering
  \small
  \caption{ANN index comparison for single-vector dense retrieval. Trade-offs between memory, latency, and recall at scale.}
  \label{tab:ann_index_comparison}
  \begin{tabular}{l|p{3cm}|p{3cm}|p{2cm}|p{5.5cm}}
  \hline
    \textbf{Index Type} & \textbf{Memory} & \textbf{Latency} & \textbf{Recall} & \textbf{Key Parameters} \\
  \hline
    Flat (Exact) & Very High & Low (10M scale) & 100\% & None (exhaustive search) \\
    IVF-Flat & Medium & Medium & 95-99\% & \texttt{nlist}, \texttt{nprobe} \\
    IVF-PQ & Low & Medium-High & 90-95\% & \texttt{nlist}, \texttt{nprobe}, \texttt{m}, \texttt{nbits} \\
    HNSW & High & Very Low & 95-99\% & \texttt{M}, \texttt{efConstruction}, \texttt{efSearch} \\
    ScaNN & Medium-High & Low & 95-99\% & \texttt{num\_leaves}, \texttt{anisotropic\_quantization\_threshold} \\
    SOAR (ScaNN+) & Medium & Very Low & 95-99\% & \texttt{redundancy\_factor}, \texttt{replication\_count} \\
  \hline
  \end{tabular}
\end{table*}

\subsection{Parameter Definitions and Tuning Guidelines}

\paragraph{IVF-Flat (Inverted File with Flat Quantization)}
\begin{itemize}
    \item \textbf{\texttt{nlist}}: Number of Voronoi cells (clusters) to partition the vector space. Typical range: 100--10,000. Higher values reduce search space per query but increase quantization error. Rule of thumb: $\sqrt{N}$ where $N$ is corpus size, or $4\sqrt{N}$ for better recall.
    \item \textbf{\texttt{nprobe}}: Number of cells to visit during search. Range: 1--100+. Higher values improve recall at cost of latency. Start with \texttt{nprobe}=1 and increase until desired recall achieved. Typical production: 8--32.
\end{itemize}

\paragraph{IVF-PQ (Inverted File with Product Quantization)}
\begin{itemize}
    \item \textbf{\texttt{nlist}}, \textbf{\texttt{nprobe}}: Same as IVF-Flat above.
    \item \textbf{\texttt{m}}: Number of subquantizers (segments to split each vector into). Must divide vector dimension evenly. Typical: 8, 16, 32, 64. Higher $m$ preserves more information but increases memory. For 768-dim vectors, $m=8$ gives 96-dim per segment.
    \item \textbf{\texttt{nbits}}: Bits per subquantizer code. Typical: 8 bits (256 centroids per subquantizer). Determines codebook size: $2^{\text{nbits}}$ centroids. Memory per vector: $m \times \text{nbits}$ bits.
\end{itemize}

\paragraph{HNSW (Hierarchical Navigable Small World)}
\begin{itemize}
    \item \textbf{\texttt{M}}: Number of bi-directional links created for each node. Range: 16--64. Higher values improve recall and reduce hops but increase memory ($\sim M \times 2 \times 4$ bytes per vector for pointers). Typical: 16--32 for production.
    \item \textbf{\texttt{efConstruction}}: Size of dynamic candidate list during index construction. Range: 100--500. Higher values create better graphs but slower indexing. Should be $\geq \texttt{efSearch}$. Typical: 200--400.
    \item \textbf{\texttt{efSearch}}: Size of dynamic candidate list during query. Range: 50--500. Directly controls recall-latency trade-off. Start with 100 and tune. Does not affect index size, only search time. Typical production: 64--128.
\end{itemize}

\paragraph{ScaNN (Scalable Nearest Neighbors - Google Research)}
\begin{itemize}
    \item \textbf{\texttt{num\_leaves}}: Number of leaf partitions in the tree. Similar to \texttt{nlist} in IVF. Typical: 1000--10,000. Determines granularity of vector space partitioning.
    \item \textbf{\texttt{anisotropic\_quantization\_threshold}}: Threshold for applying anisotropic vector quantization, which adapts quantization to data distribution. Range: 0.0--1.0. Lower values apply quantization more aggressively.
    \item \textbf{\texttt{dimensions\_per\_block}}: For product quantization, number of dimensions per quantization block. Typical: 2--8.
\end{itemize}

\paragraph{SOAR (Spilling with Orthogonality-Amplified Residuals)}
\begin{itemize}
    \item \textbf{\texttt{redundancy\_factor}}: Controls how many times each vector is replicated across partitions. Range: 1.5--5.0. Higher redundancy improves recall by allowing vectors to be found via multiple paths, at cost of index size. Typical: 2.0--3.0 for 2--3× speedup over ScaNN baseline.
    \item \textbf{\texttt{replication\_count}}: Explicit number of partition replications per vector. Alternative to \texttt{redundancy\_factor}. Typical: 2--4 partitions.
\end{itemize}

\subsection{Selection Decision Tree}

\begin{enumerate}
    \item \textbf{Corpus size < 10M vectors}: Use Flat index for exact search if memory permits.
    \item \textbf{Memory constrained (billion-scale corpora)}: Use IVF-PQ with aggressive compression ($m=8$, $\text{nbits}=8$).
    \item \textbf{Recall priority with adequate memory}: Use HNSW with $M=32$, tune \texttt{efSearch} for recall target.
    \item \textbf{Latency priority with budget}: Use ScaNN or SOAR with appropriate redundancy.
    \item \textbf{Trillion-scale (Google/Meta scale)}: Use IVF-PQ with distributed sharding, as reported by Faiss team.
\end{enumerate}






\section{Related Work and Survey}
\label{app:related_work}

This appendix provides detailed positioning of our survey relative to existing literature and explains our scope decisions.

\subsection{Comparison with Existing Surveys}

\paragraph{General Negative Sampling Surveys.}
\citet{Xu2022Review} provides a comprehensive review of negative sampling across machine learning, covering applications in recommendation systems, graph neural networks, and metric learning. However, their treatment of information retrieval is limited and predates the LLM era. Similarly, \citet{Yang2024TPAMI} surveys negative sampling for dense retrieval but focuses primarily on computer vision and cross-modal retrieval, with minimal coverage of text-only dense retrieval or modern NLP applications.

\paragraph{LLM4IR Survey.}
The LLM4IR~\cite{LLM4IRSurvey} survey offers broad coverage of how LLMs can be applied across all IR components: query rewriting, retrieval, reranking, and reading comprehension. While valuable, it treats LLM applications holistically rather than providing systematic analysis of any single training component. Specifically, LLM4IR mentions data augmentation techniques but does not: (1) categorize these through a negative sampling framework, (2) analyze the false negative problem and mitigation strategies, (3) examine computational cost vs. effectiveness trade-offs, (4) provide systematic performance comparisons, or (5) discuss technique combinations and interactions. Our survey complements LLM4IR by providing technical depth on the contrastive learning optimization process.

\paragraph{Multilingual IR Study.}
\citet{rajapakse2024negativesamplingtechniquesmultilingualsetting} compares multiple negative sampling methods specifically for multilingual IR, examining in-distribution and out-of-distribution performance. While this empirical study is valuable, it is not a comprehensive survey and focuses narrowly on cross-lingual transfer rather than the broader taxonomy of techniques.

\subsection{Complementary Approaches Not Surveyed}

Our survey focuses exclusively on negative sampling for contrastive learning in dense retrieval. Several complementary approaches improve retrieval quality but operate at different levels:

\paragraph{Knowledge Distillation.}
Training bi-encoders to mimic more powerful cross-encoder scores provides quality improvements orthogonal to negative sampling. The interaction between in-batch negatives and knowledge distillation has been studied \citep{Lin2021RepL4NLP}, but distillation itself is a teacher-student paradigm rather than a data selection strategy.

\paragraph{Inference-Time Query Augmentation.}
Methods like HyDE \citep{gao2022hyde} use LLMs to generate hypothetical documents from queries at inference time, improving retrieval without modifying training. Similarly, \citet{Lin2023Dragon} uses first-pass retrieval to rewrite queries for second-pass retrieval. These approaches enhance query representation rather than training data selection.

\paragraph{Alternative Architectures.}
ColBERT \citep{Khattab2020ColBERT} uses late interaction mechanisms, deferring fine-grained scoring to inference time. This shifts complexity away from training-time negative sampling, but ColBERT still requires contrastive training with negatives. Such architectural innovations are complementary to but distinct from negative sampling strategies.

\paragraph{Refined Representations.}
Methods like DEBATER \citep{ji2025learningeffectiverepresentationsdense} use LLM reasoning to create nuanced document representations before retrieval. These preprocessing approaches can be combined with any negative sampling technique.

\subsection{Scope Justification}

Our focused scope on negative sampling for contrastive learning serves researchers and practitioners who need to optimize this critical training component. Deep expertise in negative sampling optimization—covering random sampling, static/dynamic mining, false negative mitigation, and LLM-based synthesis—provides more value to this audience than shallow coverage of all retrieval approaches. Recent SOTA models (NV-Embed-v2, Gemini Embeddings) achieve their performance primarily through negative sampling innovations rather than architectural changes, validating the importance of our focus area.
\end{document}